\documentclass[num-refs,ngerman]{wiley-article}

\usepackage{siunitx}
\usepackage{algorithm2e}
\usepackage[utf8]{inputenc}
\usepackage{amsmath}
\usepackage{amsfonts}

\papertype{Original Article}
\paperfield{Magnetic Resonance in Medicine}

\begin{document}

\title{MRzero –- Automated discovery of MRI sequences using supervised learning}


\author[1,2\authfn{1}]{Loktyushin A}
\author[1,4]{Herz K}
\author[3]{Dang N}
\author[1]{Glang F}
\author[1]{Deshmane A}
\author[3]{Weinm\"uller S}
\author[3]{Doerfler A}
\author[2]{Sch{\"o}lkopf B}
\author[1,4]{Scheffler K}
\author[1,3\authfn{1}]{Zaiss M}

\contrib[\authfn{1}]{Equally contributing authors.}

\affil[1]{Magnetic resonance center, Max-Planck Institute for Biological Cybernetics, Tübingen, Germany}
\affil[2]{Empirical Inference, Max-Planck Institute for Intelligent Systems, Tübingen, Germany}
\affil[3]{Friedrich-Alexander Universität Erlangen-Nürnberg (FAU), Neuroradiology, University Clinic Erlangen, Erlangen, Germany}
\affil[4]{University of Tübingen, Tübingen, Germany}

\corraddress{Moritz Zaiss, Max-Planck Institute for Biological Cybernetics, Tübingen, Germany}
\corremail{moritz.zaiss@tuebingen.mpg.de}

\fundinginfo{MPG, DFG , Grant Number: 123456; ERC,Grant Number: 123459}

\runningauthor{Loktyushin et al.}

\maketitle

\begin{abstract}
\textbf{Purpose:} A supervised learning framework is proposed to automatically generate MR sequences and corresponding reconstruction based on the target contrast of interest. Combined with a flexible, task-driven cost function this allows for an efficient exploration of novel MR sequence strategies. 
\textbf{Methods:} The scanning and reconstruction process is simulated end-to-end in terms of RF events, gradient moment events in x and y, and delay times, acting on the input model spin system given in terms of proton density, T1 and T2, and $\Delta$B0. As a proof of concept, we use both conventional MR images and T1 maps as targets and optimize from scratch using the loss defined by data fidelity, SAR penalty, and scan time.  
\textbf{Results: } In a first attempt, \textit{MRzero} learns gradient and RF events from zero, and is able to generate a target image produced by a conventional gradient echo sequence. Using a neural network within the reconstruction module allows arbitrary targets to be learned successfully. Experiments could be translated to image acquisition at the real system (3T Siemens, PRISMA) and could be verified in the measurements of phantoms and a human brain \textit{in vivo}. 
\textbf{Conclusions: } Automated MR sequence generation is possible based on differentiable Bloch equation simulations and a supervised learning approach.

\end{abstract}

\section{Introduction}
Magnetic resonance (MR) images can be created non-invasively using only static and dynamic magnetic fields, and radio frequency pulses. MR imaging provides fast image acquisitions which have been clinically feasible only since the discovery of efficient MR sequences\cite{HAASE1986258,HennigRARE,CarrSSFP}, i.e., time-efficient application of two building blocks: radio frequency pulses and spatial magnetic field gradients. The proper arrangement of these building blocks is the crucial step for MR sequence development. 
In the context of MR in medicine, generating contrast between tissues is of central importance. Here MR shows outstanding properties especially in soft tissues, leading to many applications in routine medical imaging with specialized MR sequences for a certain contrast of interest. The direct relationship between MR image contrast and the actual MR sequence with its many free parameters raises the question if both image and contrast generation can be performed in a completely automatic manner. Many recent works addressed this task by analytic optimization of RF pulses and image acquisition parameters \cite{roux1993stabilization, shinnar1989synthesis, hargreaves2004time, lustig2008fast}. 
\\A principal idea different from previous approaches would be to use a certain desired image contrast as a target, and then automatically create the MR sequence and signal-to-image reconstruction routine that is able to generate this contrast using advanced optimization techniques. The idea of jointly learning acquisition and reconstruction of MR images received  a lot of attention recently. Most of this work deals with the problem of accelerated imaging, and attempts to learn optimal k-space undersampling patterns that minimize scan time, while still obeying scanner limitations and allowing for artifact-free reconstruction \cite{jin2019selfsupervised,weiss2019learning,sherry2019learning,bahadir2019learningbased,weiss2019pilot}. These approaches already link acquisition and reconstruction optimization, although the mentioned approaches only deal with the sampling pattern aspect of MR measurements and don't optimize for pulses, gradient forms and timing.
Novel approaches address the problem of general sequence learning\cite{Zhu_autoseq2018,walker-samuel_2019,Shin_et_al_2019}. The authors propose using deep reinforcement learning to find the optimal sequence satisfying a given loss function. A first approach of automatic sequence generation (AUTOSEQ) for 1D imaging was presented by Zhu et al. based on Bayesian deep reinforcement learning \cite{Zhu_autoseq2018}. While deep reinforcement learning is highly flexible, it is also costly in terms of computation time, as many parameter paths must be explored before the best path can be found. In contrast, using a gradient based method makes it possible to directly follow the locally steepest gradient path to minimize the loss function. Such an approach was also shown in another work of Zhu et al. using auto-differentiation \cite{Zhu_autoseq2019}. In the first approach, 1D images could be generated based on the classic Fourier transform. In the second approach, only single pixel (0D) evaluations were shown, but with full quantification of the Bloch parameters. Another recent sequence generation approach with a completely different target formulation was introduced by Walker-Samuel \cite{walker-samuel_2019}. Here, sequence parameters were also updated using deep reinforcement learning but the task was instead of generating an image directly, the task was to guess a geometric shape in the scanner. Interestingly, by solving this simple task, the agent was then able to generate an edge detector. 
\\
In the present work we propose a joint automatic sequence and reconstruction optimization framework, \textit{MRzero}, based on supervised learning and a fully differentiable MR simulation process. This enables gradient descent in the sequence parameter space, as well as efficient generalization using multiple input and target samples. We show how basic image encoding as well as artifact suppression and SAR minimization can be achieved. Furthermore, we show how arbitrary target images can be used when using a neural network layer added to the reconstruction, and how possible overfitting can be tackled in an efficient manner. 
\textit{MRzero} performs the optimization in a simulation environment -- a differentiable simulated replica of the MR system. However, the learned sequences were implemented and tested on a real clinical MR system and scanned both \textit{in vitro} and \textit{in vivo}. The code for our implementation is available upon a collaboration request.

\section{Theory}

In this work we address the problem of automated sequence design and optimization. The approach that we propose allows optimization over spatial encoding gradients (2D), excitation flip angles and phases, and timing of different events within a sequence affecting the relaxation weighting. The optimization is carried out in an MR scanner simulation environment (implemented in Torch \cite{paszke2017automatic}) mirroring the acquisition of a real MR scanner (see also Figure 1). Mathematically, the forward simulation amounts to a chain of tensor-tensor multiplication operations that are applied to a tensor $\mathbf{m} \in \mathbb{R}^{N_{vox} \times N_{spins}\times 3}$ describing a current magnetization state (magnetization is a 3-component vector). The $N_{vox}$ and $N_{spins}$ constants determine the number of voxels and isochromats per voxel, respectively, used for simulation. The forward process is differentiable in all parameters, and supports an efficient analytic derivative-driven non-linear optimization, which is guaranteed to converge to a local optimum. Furthermore, since the simulation in each isochromat is independent of all other isochromats, we can parallelize the computations over GPU threads with a minimal implementation overhead, relying on efficient libraries such as Torch and Tensorflow\cite{tensorflow2015}, which, although designed for Deep Learning, are well-suited for our purpose. \\
Figure 2 outlines the general method pipeline and problem definition. The pipeline consists of two modules: acquisition and reconstruction. The inputs to the pipeline are the spin system characterization variables of proton density (PD), relaxation times (T1 and T2) and off-resonance frequency (B0), and the output is the image. The proposed framework makes it possible to formulate and solve the following optimization tasks: \\
1. Given the target image/contrast of interest and associated spin system descriptor (PD/T1/T2/B0 defined at each voxel) find optimal sequence parameters that produce simulation results as close as possible to the target in the L2 norm sense. \\
2. Augment the task in 1 with additional constraint terms that allow a sequence with certain aspects of interest. For example, prefer low SAR, shorter scan times, or higher SNR. \\
3. Given multiple targets/spin system descriptor inputs, find a sequence that implements tasks 1 and 2 generalizing over data statistics of the input/target pairs. Such a task closely resembles a supervised learning regression problem, where a scanner function of sequence parameters is used as a trainable model. Optionally, a neural network can be concatenated to a scanner function, and its weights learned simultaneously, allowing us to train a complete measurement-reconstruction chain.

\subsection{Model description}
The forward model is discretized not in time but as a chain of subsequent Bloch events, each consisting of an RF event defined by the the flip angle $\boldsymbol{\alpha}$ and phase, a gradient precession event defined by the gradient moments $\mathbf{g}$, a free precession event and a relaxation event with the time variable $\boldsymbol{\Delta}\mathbf{t}$, and a binary mask if the ADC is open or closed $\mathbf{d}$.

The forward process of the image acquisition that we use is outlined in an Algorithm box below. For more details on operator definitions, see Supporting Information Figure S1.

\begin{small}
\begin{algorithm}[H]
\vspace{2 mm}
  Algorithm 1: Initialize: $\mathbf{m}$ to initial magnetization state: $\mathbf{m} = \mathbf{m}_0 \propto \hat{\mathbf{e}}_z \cdot PD$, where $PD \in \mathbb{R}^N_{vox}$ is a proton density vector \\
    \vspace{1 mm}
  \For{$r \displaystyle \leftarrow 0:N_{rep}$}
  {
  \vspace{1 mm}
  \For{$a \displaystyle \leftarrow 0:N_{events}$}
  {
    \vspace{1 mm}
    $\mathbf{m} = FLIP_y(\alpha_{r,a}\cdot B1^{+}, \phi_{r,a})\mathbf{m}$ \\
    \vspace{1 mm}
    $\mathbf{m} = RELAX(\Delta t_{r,a})\mathbf{m} + (1-RELAX(\Delta t_{r,a})) \mathbf{m}_0$ \\
    \vspace{1 mm}
    $\mathbf{m} = FREEPRECESS(\Delta t_{r,a})\mathbf{m}$ \\
    \vspace{1 mm}
    $\mathbf{m} = GRADPRECESS(g_{x_{r,a}},g_{y_{r,a}})\mathbf{m}$ \\
    \vspace{1 mm}
    $s_{r,a} = ADC(d_{r,a})\mathbf{m}_{transverse} + \bm{\varepsilon}$ \\
  }
  }
    \vspace{1 mm}
  Output: signal $\mathbf{s} \in \mathbb{R}^{N_{rep}\times N_{events}\times 2}$ (complex-valued but stored in a real number type array)
 
\end{algorithm}
\end{small}

Given that each \textit{event} operator in the chain is linear we can concatenate all tensors into a single
linear operator $\textit{SCANNER}$, representing a scanner equation in a compact form:

\begin{equation} \label{eq:SCANNERfunction}
\mathbf{s} = \textit{SCANNER}(\boldsymbol{\alpha},\boldsymbol{\Delta} \mathbf{t},\mathbf{g},\mathbf{d})\mathbf{m}_0,
\end{equation} where $\mathbf{m}_0 = \hat{\mathbf{e}}_z \cdot PD$ is an initial magnetization state scaled by the proton density at each voxel, and $\boldsymbol{\alpha} \in \mathbb{C}^{N_{events} \times N_{rep}}, \boldsymbol{\Delta} \mathbf{t} \in \mathbb{R}^{N_{events} \times N_{rep}}, \mathbf{g} \in \mathbb{R}^{N_{events} \times N_{rep} \times 2}, \mathbf{d} \in \{0,1\}^{N_{events}}$. Although at each step of the algorithm chain only linear transformations are applied to magnetization tensor, the \textit{event} tensors themselves are composed of elements that depend in a non-linear way on both sequence parameters and spin system characterization parameters such as T1 and T2. \\
In our simulation, we distinguish two types of input variables:

1. Scanned object descriptor variables: $PD, T1, T2, R2^{*}, B0 \in \mathbb{R}^{N_{vox}}$. These parameters are fixed during optimization, and are given as input to the algorithm for a given object sample.

2. Scanner sequence variables: $(\boldsymbol{\alpha},\boldsymbol{\Delta} \mathbf{t},\mathbf{g},\mathbf{d})$. These parameters influence the signal $\mathbf{s}$ output by the scanner operator, and can vary during the optimization process.

Although our model allows for a simple extension to a multi-channel receive/transmit coil setup and parallel imaging, in the current proof-of-principle work we use a single receive element in both simulations and real measurements. We also assume uniform receive sensitivity of this element, which leads to the fact that $PD$ used in the simulation corresponds to $PD$ times the receive sensitivity map. 

\subsection{Reconstruction}
In case the of Cartesian acquisition, image reconstruction can be performed by applying a Hermitian transpose of an orthonormal Fourier transform matrix to the measured signal: $\mathbf{r} = \mathbf{F}^{H}\mathbf{s}$. In the more general case of arbitrary k-space trajectories, it is beneficial to consider the following proxy acquisition equation (we omit relaxation, free precession and excitation terms here for the ease of presentation):

\begin{equation} 
{\displaystyle s_{r,a}=d_{r,a}\sum_{N_{vox},N_{spins}}\mathbf{m}_{transverse}(x,y)exp(i(\sum_{a=0}^{N_{events}}(g_{x_{r,a}})x+\sum_{a=0}^{N_{events}}(g_{y_{r,a}})y))}
\end{equation} 
where $r$ and $a$ are repetition and event indices. We can rewrite the equation in matrix form:  $\boldsymbol{s}=\mathbf{E}(\mathbf{g})\boldsymbol{m}_{transverse}$, where $\mathbf{g}$ are the gradient moments (integrated gradient events over time) within each event, and $\mathbf{m}_{transverse}$ is a 2-component vector describing magnetization state in the transverse plane at each voxel before encoding gradients are applied. In the case of Cartesian encoding, the matrix $\mathbf{E}$ is equal to a discrete Fourier transform matrix $\mathbf{F}$. 
In the case of general encoding, we approximate the reconstructed image as $\mathbf{r} \approx \mathbf{E}(\mathbf{g})^{H}\boldsymbol{s}$. When the matrix $\mathbf{E}$ is orthonormal, the inverse of $\mathbf{E}$ is equal to its Hermitian and the reconstruction is exact. \\
Using an adjoint of an encoding operator $\mathbf{E}$ makes it possible to avoid using non-uniform Fourier transforms, which require gridding and density compensation. The adjoint of the encoding matrix is differentiable and straightforward to implement. As we show later in the results section, it is also robust to small deviations from uniform k-space sampling.
For larger deviations, as in radial or spiral imaging, applying the adjoint of the encoding matrix on reconstruction would result in blurring when the k-space center is oversampled. To resolve this we use an iterative adjoint approach (see the Supporting Information section 'radial MR').
\\
As elaborated below, the output of the adjoint operator can be the input to a subsequent neural network, which we use in one example to generate the final output image. We refer to this form of reconstruction in Figure 1 as reconstruction type II.

\section{Methods}

\subsection{Target fitting/supervised learning}
Let's assume that the target image $\mathbf{r}$ that we want to fit is given. In case we only optimize for gradient moments we can formulate the following objective function:

\begin{equation} 
\displaystyle \mathbf{g}^*=\underset{\mathbf{g}}{\arg\min}(\left\Vert  \mathbf{r} - \mathbf{E}(\mathbf{g})^{H}\mathbf{E}(\mathbf{g})\mathbf{m}_{transverse}) \right\Vert ^{2}_2),
\end{equation} 
or, in case we optimize for all sequence parameters, using Eq. \ref{eq:SCANNERfunction} :

\begin{equation} 
\displaystyle \boldsymbol{\Psi}^*=\underset{\boldsymbol{\Psi}}{\arg\min}(\left\Vert \mathbf{r} - \mathbf{E}(\mathbf{g})^{H}(\textit{SCANNER}(\boldsymbol{\Psi},\mathbf{m}_0)) \right\Vert ^{2}_2)
\end{equation} 
where $\{\boldsymbol{\alpha},\boldsymbol{\Delta} \mathbf{t},\mathbf{g}_{x,y}, \mathbf{d}\} = \boldsymbol{\Psi}$ are sequence parameters.

Additionally, we augment the objective function with extra terms that put a penalty on optimized parameters of choice. For example, if we wish to find a sequence that keeps a moderate SAR, the following objective can be used:

\begin{equation} 
\displaystyle \boldsymbol{\Psi}^*=\underset{\boldsymbol{\Psi}}{\arg\min}(\left\Vert \mathbf{r} - \mathbf{E}(\mathbf{g})^{H}(\textit{SCANNER}(\boldsymbol{\Psi},\mathbf{m}_0)) \right\Vert ^{2}_2 + \lambda\left\Vert \bm{\alpha} \right\Vert ^{2}_2)
\end{equation} 
where the extra term puts an L2 penalty on the magnitude of RF pulses. The $\lambda$ hyperparameter balances the error made in reconstructing the image and SAR reduction level.

The proposed objective function is non-linear in sequence parameters, since the optimized variables are the arguments of $sin$,$cos$ and $exp$ functions that form the elements of the tensors from Algorithm 1. Non-linearity of the objective implies that the loss is non-convex and therefore we are not guaranteed to reach an optimal solution. In fact, depending on initialization and resolution of the image, the obtained solution can be arbitrarily poor. To cope with this problem we resort to performing random restarts of the optimizer with resets of the running optimization step size. It is important to note at this point that although we use the term optimization ubiquitously throughout the paper, the actual goal of this work is not to reach a certain pre-defined "optimality" condition (i.e. being better than existing sequences, or finding a "perfect" solution to a certain problem) but rather to drive the learned sequence towards a solution that approximates the target of interest.
Optimization is performed using the optimizer Adam\cite{kingma2014adam}, with the step parameter varied from high to low values over the course of the optimization to promote exploration in the beginning and convergence and stability at the end. Normally, around 1000 iterations are required to converge to result with less than $10\%$ NRMSE (Normalized Root Mean Square Error).

\subsubsection{Supervised learning with a sparse synthetic training set}
To avoid overfitting, we can define an entire set of target/input pairs, and try to learn a scanner function that minimizes the combined error over all training pairs:

\begin{equation} 
\displaystyle \boldsymbol{\Psi}^*=\underset{\boldsymbol{\Psi}}{\arg\min}(\sum_{q}^{N_{samples}}\left\Vert \mathbf{r}_q - \mathbf{E}(\mathbf{g})^{H}(\textit{SCANNER}(\boldsymbol{\Psi},\mathbf{m}_{0,q})) \right\Vert ^{2}_2 + \lambda\left\Vert \bm{\alpha} \right\Vert ^{2}_2),
\end{equation} 
where $N_{samples}$ is the number of input/target pairs in the training set. 
In this research work, we generate a synthetic database of input/target pairs, where each sample in the dataset represents a scanned object primitive. We define such primitives to be just single voxels of non-zero PD at varying spatial locations (see Figures 6 and 7). Additionally, we vary the T1/T2/B0 value of each voxel primitive in the dataset. We then make a forward pass subject to a sequence of interest to produce target images that we want to learn to output. A single voxel approach has a significant computational benefit, since we only need to perform Bloch simulations for isochromats within a single voxel (see also Supporting Information Figure S2).

\subsubsection{Neural network extension to reconstruction module}

We extend the versatility of the reconstruction module by augmenting it with a neural network. As an input to the neural network we use the output of the adjoint operator. Stacking the neural network at the end of the forward process chain makes it possible to model diverse targets such as binary segmentation masks or T1 and T2 maps:

\begin{equation} 
\displaystyle \boldsymbol{\Psi}^*,\boldsymbol{\Theta}^*=\underset{\boldsymbol{\Psi},\boldsymbol{\Theta}}{\arg\min}(\sum_{q}^{N_{samples}}\left\Vert \mathbf{r}_q - NN_{\boldsymbol{\Theta}}\Big(\mathbf{E}(\mathbf{g})^{H}(\textit{SCANNER}(\boldsymbol{\Psi},\mathbf{m}_{0,q}))\Big) \right\Vert ^{2}_2 + \lambda\left\Vert \bm{\alpha} \right\Vert ^{2}_2)
\end{equation} 
the neural network weights $\boldsymbol{\Theta}$ are learned alongside with sequence parameter optimization.  To avoid the problem of overfitting and reduce complexity, purely convolutional neural network architectures can be employed. The first layer of the network can have either two channels (real and imaginary output from adjoint), or non-combined output from the adjoint, such as the number of channels equal to $2N_{rep}$.

\subsection{Implementation and computation time}

Our forward process can be efficiently parallelized with the use of GPUs. The parallelization is carried out over isochromats and voxels which can be treated independently. Although, the forward process is then efficient to carry out, the simulation can be lengthy due to multiple repetitions and samples per readout. This is because, tensor-tensor operations need to be performed sequentially, when proceeding through a chain of events and repetitions. 
Our current implementation uses 625 isochromats per voxel; the large number of isochromats is required to properly describe transient net magnetization, especially in case of spoiled gradient echo-like sequences. 
Already for 48x48 voxels this implies more than a million of variables that need to be manipulated at every \textit{event} step, which gives our  model a high computational cost. Depending on the task, computation time is certainly a limitation of the approach in the current implementation. A typical optimization took \textasciitilde 7 hours until convergence at \textasciitilde 25s per iteration and approximately 1000 iterations in total on Nvidia V100 GPU. The T1-auto-encoder task presented in Results section, with 6 measurements at 64x64 resolution took \textasciitilde 30 s per iteration and 500 iterations in total. Supporting Information Figure S2 shows a computation time analysis that reveals the quadratic increase in run time with resolution and number of isochromats, as well as the decrease in run time when using a smaller block through the sparse synthetic approach. 

In all of the experiments that we present in the current work we assume a fixed number of repetitions ($N_{rep}$) and events ($N_{events}$). Here, $N_{rep}$ is set to the matrix size, and $N_{events}$ to the matrix size + 4 to allow two events before and after the ADC. Furthermore, we assume a hard-coded definition of parameters which can be optimized at each event index. For example, for a RARE sequence we always assume the first two events are dedicated to RF, and subsequent events to spatial encoding and rewinding. Supporting Information Figure S1 illustrates a typical matrix of repetitions/events.

\subsection{Data acquisition at a real MR system}
After the full in-silico optimization process, certain intermediate iterations were executed on the real system in phantoms and in vivo. To run sequences on a real MR system the sequence parameters were exported using the pulseq standard  (\cite{layton_pulseq_2017}, pulseq.github.io). Pulseq files could then be interpreted on an MRI scanner including all necessary safety checks, and were executed on a PRISMA 3T scanner (Siemens Healthineers, Erlangen, Germany) using a 2 and a 24 channel Head Coil. Raw data was automatically sent back to the terminal and the same reconstruction pipeline used for the simulated data was used for measured image reconstruction. During export to the pulseq format, actual gradient events with finite slew rates and amplitude were calculated to realize the demanded gradient moment. To achieve that we introduced minimal event times that are long enough to generate  gradients with moments in the order of 1.5 \textit{k}-space span. The same minimal times were added to the simulation such that it was not the total event time that was optimized but rather the strictly positive variable that was added to the minimum event time. To be able to run slice selective experiments, block pulses were substituted during export by 1 ms sinc-shaped pulses with time-bandwidth product of 4, and corresponding slice selection and rephasing gradients. 

For phantom scans the resolution phantom of Bruker was positioned at the iso-center. This phantom was fully characterized for a 50 Hz shim using T1, T2, B1 and B0 mapping to generate the exact same phantom in silico. For proton density (PD) a fully relaxed centric-reordered FLASH \cite{HAASE1986258} image was used, thus the coil sensitivities are actually modelled as PD variation in the simulation.   
To avoid sequence abortion by the scanner due to exceeding SAR limits and potential peripheral nerve stimulation, we used rather conservative building blocks: RF pulses were always contained in a 2 ms block, normally with 1 ms RF duration. Sampling interval in read direction was 0.08 ms corresponding to a bandwidth of 195 Hz/px for a 64-pixel image matrix width. In vivo human measurements were approved by the local ethics board.

\subsection{Experiments / Learning tasks}
This work demonstrates three different optimization tasks: 
\begin{itemize}
\item \textbf{\textit{Encoding and RF learning task}} (Figures 3, 4, 5a,c, and 8): learning of encoding from zero using a GRE image as a target. Here, both RF and gradient moment events were optimized. In Figure 4 the event times were also optimized.
\item \textbf{\textit{RF optimization task.}}: experiments on the importance and effect of SAR regularization of excitation pulses using a GRE target image (Figure 5b). Event times and read gradients were fixed. Figures 6 and 7 consider RARE RF optimization upon SAR regularization. With 90$^{\circ}$ excitation and event times and gradients fixed, refocusing RF pulses were optimized.
\item \textbf{\textit{Quantitative T1 auto-encoding task.}} (Figure 9): T1 auto-encoding using a neural network in the reconstruction (path II in Figure 1). Here, the sequence timing and neural network are optimized simultaneously. RF events and gradients were fixed.
\end{itemize}
Both the \textit{encoding and RF learning task}, and the \textit{RF optimization} were trained and evaluated on various datasets: phantom image, brain image and sparse synthetic dataset.

\section{Results}
\subsection{Encoding and RF learning task}
In the first example, the target is a Cartesian gradient echo image with constant flip angle of 5 degree, as visible in column 6 of Figure 3. A sequence with 96 acquisitions is initialized with all gradients and flip angles set to zero. An additional SAR penalty was added to the cost function, explained in more detail in the next paragraph. The initial state is still visible in the first iterations (column 1 in Figure 3) showing all line acquisitions collapsed starting from $k = 0$ (Figure 3a) and small flip angles (Figure 3b), leading to a non-encoded image in simulation and real measurements (Figure 3cde). 
 During the training, the image error with respect to the target decays from 1000 \% to 10 \% (Figure 3f) by inventing suitable x- and y-gradient events, as well as non-zero RF amplitudes and phases. The achieved error is below randomly reordered Cartesian sequences (Supporting Figure S8). This enables gradual improvement of the image encoding and the structure of the phantom becomes visible after 300 iterations, converging at around 600 iterations. The obtained sequence is non-Cartesian and exhibits a complex RF pattern. Despite this, there is a close match with the target image in the simulation (Figure 3c) and in the real measurement of the corresponding phantom (Figure 3d). The high coherence between our simulation and the measurement is depicted by very similar intermediate images. Interestingly, the sequence generalizes well to an \textit{in vivo} measurement of the human brain at 3T. Image errors with regard to the in vivo target were also calculated, but are actually close to the simulation image errors and therefore not shown. Please note that the target image does not have to be necessarily an MR image; a binary mask can also be used as a target as shown in Supporting Information Figure S3. In Figure 4, event times and read gradients were additionally optimized from zero. This is the most complicated task with most degrees of freedom and has lower convergence and k-space coverage, but the target-based optimization is able to generate a functional sequence with different TE an TR for every repetition and even inverted readout direction for some repetitions while maintaining an image error below 20\%.
 Thus, Figures 3 and 4 show already demonstrate the potential of \textit{MRzero}: the invention of a complete MRI sequence that is applicable to image acquisition at a real 3 T system in phantoms and \textit{in vivo}.
 
 In Figure 5 we show the importance of SAR regularization. Figure 5a-d shows the previous experiment, but without any SAR penalty (at lower resolution). In this case the optimizer can choose very high flip angles, e.g. multiples of 180 and 360 degrees, to generate the target image. Figure 5e-h shows the outcome of a sequence initialized to the target sequence and optimized only for RF events (this reflects a pure \textit{RF optimization task}) upon a SAR penalty. The optimizer lowers flip angles in the outer k-space lines to reduce SAR by 27 \%, tolerating a small image error. A SAR penalty added to the cost function regularizes the flip angles and gradients, and leads to a low flip angle solution shown in Figure 5i-l. 
 \subsection{RF optimization task}
While a SAR restriction is normally not necessary for a low-angle GRE sequence, the same approach can be performed for the more SAR intensive RARE sequence shown in Figure 6. Here the refocusing flip angles were optimized using \textit{MRzero} under a strong SAR penalty. In the numerical brain phantom the RARE sequence shows discernible T2-weighting; given the target, \textit{MRzero} tries to reduce SAR while maintaining the contrast. We observe that SAR can be reduced to ~15 \% of the target sequence value without compromising much of image quality or contrast. In later iterations the strong SAR penalty leads to high signal intensities and blurry images. 
Despite the observed generalization to \textit{in vivo} applications, these approaches are prone to overfitting as only one phantom was used during the training with limited distribution of T1 and T2. This can be seen when using the disc phantom for training, as done in Supporting Information Figure S3, where very different flip angle trains are found by \textit{MRzero}. To overcome such possible overfitting biases we suggest using a suitable simulation for the organ of interest, or to use the more general approach described in the following section.

\subsection{Supervised learning with a sparse synthetic dataset approach}
To overcome overfitting, we have employed the sparse synthetic dataset approach (see Methods section). A large database (1024 samples) of input and target blocks at different positions and with different MR properties was generated and used in the training phase in a randomized manner. At every optimizer step, a minibatch containing a single sample randomly drawn from the database was used to compute the loss and update the parameters. Directly applied to the previous SAR restricted RARE, Figure 7 shows the outcome of this approach, dubbed sparse synthetic training set learning. Although the training set only consists of very sparse data (Figure 7b), a sequence scheme (Figure 7a) similar to the case of brain phantom training (Figure 6a) is found, and again the contrast is maintained down to SAR values of 10\%. This SAR restricted RARE optimization was solved before conventionally \cite{HennigTRAPS2003, sbrizzi2017optimal, busse2006fast}, and the flip angle variation MRzero found compares well to these previous approaches.  Note, that we always show a different randomly chosen sample at every iteration in the figure, thus the apparent change of the sparse voxel position. Furthermore, we show just a single sample from the database as a target; in practice there is a unique target for each of the 1024 examples.
With such a sparse synthetic training approach, spatial encoding strategies can be learned, which brings us back to our initial example of Figure 3.  

Figure 8 again shows the experiment presented in Figure 3, using sparse synthetic data for training. It shows that the spatial encoding can also be learned by moving a small block over the FOV during the training iterations. Thus, learning on a sparse synthetic dataset can generally be applied to avoid overfitting. Again, generalization of the learned sequence to both phantom and \textit{in vivo} measurement could be validated (Figure 8de).

\subsection{Quantitative T1 auto-encoding task}
In a final experiment we performed a joint optimization of sequence and weights of a neural network in the reconstruction module to generate a T1 mapping sequence (case II in Figure 1). All details of this experiment can be found in the Supporting Information. To reduce parameters we concatenated six inversion-prepared gradient echo readouts, and set the recovery time Trec after each readout and inversion time TI after each inversion pulse to 0 (Figure 9a). Trec and TI were then learned jointly with the pixel-wise neural net with a synthetic dataset approach and T1 values as a target. After several iterations, reasonable Trec and TI times were found (Figure 9ab), and the prediction matched literature values within the standard deviation, as can be seen in Figure 9cd (in vivo data from a 3T brain measurement). 

\section{Discussion}

\subsection{General summary}
Herein we present a framework for automatic MR sequence generation using arbitrary target image data. We showed that basic sequence learning is possible from zero without using existing knowledge of human MRI experts, based only on the underlying laws given by the Bloch equations.
The major finding of this work is that such an approach, which takes both signal encoding and image reconstruction  steps and the many parameters into account, is actually functional and can be translated from simulation to real measurements. To validate this we made several comprehensible experiments, and in all cases the functionality of the method could be shown. 
In the following we discuss this advancement in the context of literature: the studied tasks, the outcome of these tasks, and the optimization approaches presented. Recent learning-based approaches similar to the present work are discussed as well. 

The first task of learning image encoding from zero is most similar to general work on k-space trajectory optimization for which many solutions exist \cite{pruessmann2001advances}. Seeger et.al. \cite{seeger2010optimization} formulate optimization of k-space sampling for nonlinear sparse MRI reconstruction as a Bayesian experimental design problem, where statistical sparsity characteristics of images are incorporated by way of a prior distribution. More recent approaches such as Jin et.al. \cite{jin2019selfsupervised} simultaneously train two neural networks, the first dedicated to reconstruction and the second to learning the sampling, both being linked with an automatically generated supervision signal created with Monte Carlo tree search. Weiss et.al. \cite{weiss2019learning} propose a fully differentiable end-to-end deep learning-based paradigm to learn an undersampling mask and the reconstructed image jointly. Sherry et.al. \cite{sherry2019learning} use a supervised learning approach to find the optimal sparse sampling pattern that balances acquisition time and the image quality. Until now our approach cannot compete with these efficient approaches, as it does not aim for undersampling yet, but the principle idea of selecting -- or in our case adjusting -- the gradients for a certain k-space sampling is the same. However, in contrast to previous work, the altered acquisition pattern is directly simulated in MRzero and the effect of the signal decay properly incorporated due to the full simulation. This enables the adjustment of associated flip angles for each k-space acquisition, which is helpful for certain acquisition patterns as shown  for radial, as well as free acquisition examples in Supporting Information Figures S5 and S6. 
Hand in hand with k-space sampling goes image reconstruction, for which we employed a differentiable adjoint formalism. Using the adjoint formalism has the important benefit that no regridding is needed but arbitrary k-space sampling is still possible. For k-space sampling with non-uniform density, such as radial imaging, a generalized adjoint approach (see Supporting Information Figure S5) can be used to perform differentiable reconstruction. 
In future work, using neural networks and representative quantitative MR image databases, we expect the priors on real image distribution to be implicitly encoded in the weights of the neural network \cite{knoll2019deep} leading to a further improvement in the quality of reconstructed images and faster measurement times. The observed remaining image errors e.g. in Figure 4, can most probably also be resolved using iterative or deep learning reconstruction as used for undersampled data.

The adjoint operator  fulfills the criteria of a known operator \cite{maier2019learning}, which was shown to yield theoretically smaller errors compared to a general neural network. In future work, the adjoint formalism can possibly be substituted by a neural network such as AUTOMAP \cite{zhu2018image} and still jointly optimized. Thus, combination of the adjoint formalism with a post-processing neural network module makes it possible to use prior knowledge and avoid demanding fully connected networks. Again, previous conventional and recent deep learning based approaches for optimization of encoding are compatible with the MRzero framework, and from our results we can state that investigation of hybrid approaches with a supervised training is promising. 

The second task of reducing the SAR of RARE sequences is an old problem that was solved previously in an analytic manner in several works by requiring the non-encoded echo-train intensity to be constant \cite{HennigTRAPS2003, sbrizzi2017optimal} or an unaffected point-spread function \cite{busse2006fast}. The principle outcome of our target-based supervised approach, with a refocusing flip angle around 120 $^{\circ}$ and a continuous decrease to lower flip angles, matches these previous works and could validate our method. Importantly, here the sparse synthetic database approach was important to avoid over-fitting (compare Supporting Information Figure S4 and Figure 6). The over-fitting here originated from the correlation of the T1 and T2 times in the disc phantom, thus an over-fitted solution could be found that made use of this correlation but did not generalize to the brain phantom. We can also relate our approach to optimal control techniques, often used for RF pulse shape design \cite{conolly1986optimal, vinding2012fast, lapert2012exploring, rund2017magnetic}. While pulse shapes were not yet considered herein, adding them to the model would be a logical next step for future extensions. 

The third task of direct mapping of quantitative MR properties falls into the category of deep MR fingerprinting. In our quantification experiment (Figure 9) focused only on T1 mapping, we showed that \textit{MRzero} can be treated as a complete auto-encoder and in principle novel MRF paradigms can be explored and optimized autonomously. Previously, many detailed analyses were performed for MRF schedule optimization \cite{cohen2017algorithm}. The first joint optimization of sequence and reconstruction using neural networks was shown for 0D by Zhu et al \cite{Zhu_autoseq2019}. \textit{MRzero} is able to perform such joint optimization in 2D, following the same approach as shown for T1 mapping, validated by real measurements. This is a proof-of-concept that can be extended to multi-parametric mapping - similar to MR fingerprinting - yielding proton density, T1, and T2, as well as B1 and B0 inhomogeneity maps. While these contrasts could directly be used as multi-parametric targets with the current implementation, for advanced MRI contrast like diffusion- or chemical exchange-weighted MRI the forward Bloch model can be extended for Bloch-Torrey or Bloch-McConnell models, respectively. Current developments in magnetic resonance fingerprinting (MRF)   \cite{virtue2017better,ma_MRF_2013,poorman_magnetic_nodate,mcgivney_magnetic_nodate,korzdorfer_reproducibility_2019} will as well lead to larger quantitative databases for different organs and field strengths, which are then again available for improving simulation based training.
Before MR fingerprinting approaches, such a joint optimization for T1 mapping was solved analytically as well  \cite{teixeira2018joint}. Here, the authors aimed to increase the precision of single-compartment DESPOT relaxometry by a joint system relaxometry approach that estimates parameters in a single step using all available data and optimizing acquisition parameters based on the Crámer-Rao lower bound.

\subsection{Supervised learning in the context of other optimization approaches}
Optimization of MR sequences is a well-studied problem and the major novelty of this work is to establish a very general framework for optimization, but solely using the target contrast of interest directly as a problem definition. 
In more formal language, sequence generation is formulated as a supervised learning problem, where the term "supervised learning" refers to the process of learning a function that maps an input to an output based on example input-output pairs. These pairs are given by (i) the signals of voxels with certain MR properties, generated and encoded using the end-to-end simulation, and (ii) the corresponding target voxel signal. Most of the cases presented in this work are pseudo-supervised (\textit{curve fitting} is a more appropriate term) as e.g. a phantom with only a few different T1 and T2 values was used. Thus only a few different input and output pairs were available during training. Such small training databases are in general known to lead to overfitting during the training, which means that the learned sequences can only be faithful to T1 T2 and B0 distributions similar to those used in simulations. Interestingly, for image encoding in GRE-like sequences (Figure 3-5), we did not observe overfitting to be a problem. However, when it comes to flip angle amplitudes in RARE sequences, which have more effect on the actual contrast, overfitting was clearly observed when using different sequence strategies in the simple phantom compared to the brain phantom (see Figure 6 and Supporting Information Figure S4). We addressed overfitting issues by proposing a sparse synthetic training set approach which was composed of single pixels sampled from a database that had similar PD/T1/T2 distributions as the human brain. This approach has the advantage of generality and computational efficiency.

 As explained in the Methods section, computation times are long and impose resolution constraints on our approach. However, these may be relaxed in the future by improvements in software and hardware. A more general drawback of a target-learning-based approach is that many different examples must be learned to achieve generalization. Such multi-example training is of course slower in convergence and in general more difficult to optimize compared to existing analytic approaches for sub-problems such as RF optimization. By introduction of the sparse synthetic data supervised learning approach we tried to improve both computation times and generate general and diverse data. The improvement in computation time using the sparse synthetic voxels can be seen in Supporting Information Figure S2. 
 
 While we focused on the principal framework and several comprehensive experiments herein, having such a differentiable MRI process at hand paves the way to a novel way of generating an MR sequence and reconstruction solely based on the target provided, which can be a certain MR contrast, but the possibilities for targets are limitless, e.g. quantification, segmentation, as well as contrasts of other imaging modalities.
 
\section{Conclusion}
We have developed a fully automated MRI sequence generator based on the Bloch equation simulations and supervised learning. We were able to achieve close agreement between simulated results and real measurements and demonstrated functionality of the proposed target-driven sequence inference. The approach was verified in several experiments in which the functionality of the method was demonstrated. The differentiable Bloch simulation of many samples is computationally demanding, but manageable, and the flexibility and generality of the approach opens many opportunities.   

\section*{Acknowledgements}
We thank Maxim Zaitsev and his team for developing and sharing pulseq \cite{layton_pulseq_2017}. The research that we have done would not be possible without this versatile and powerful framework. We also thank the authors and developers of pypulseq
 library \cite{ravi2018pulseq}, for the parts which we used to build a python interface to pulseq. \\
 \\
 
\hspace{-6mm} Funders: \\
Deutsche Forschungsgemeinschaft	DFG SCHE 658/12 \\
\hspace{2mm} ERC Advanced Grant	No 834940 \\
\hspace{2mm} European  Union’s  Horizon  2020  research  and  innovation  program 	667510 \\
\hspace{2mm} German  Research   Foundation 	DFG,   ZA   814/2-1

\bibliography{sample}
\newpage

\listoffigures

\section*{Figures and Captions}

\begin{figure}[!ht]
\centering
\includegraphics[width=14cm]{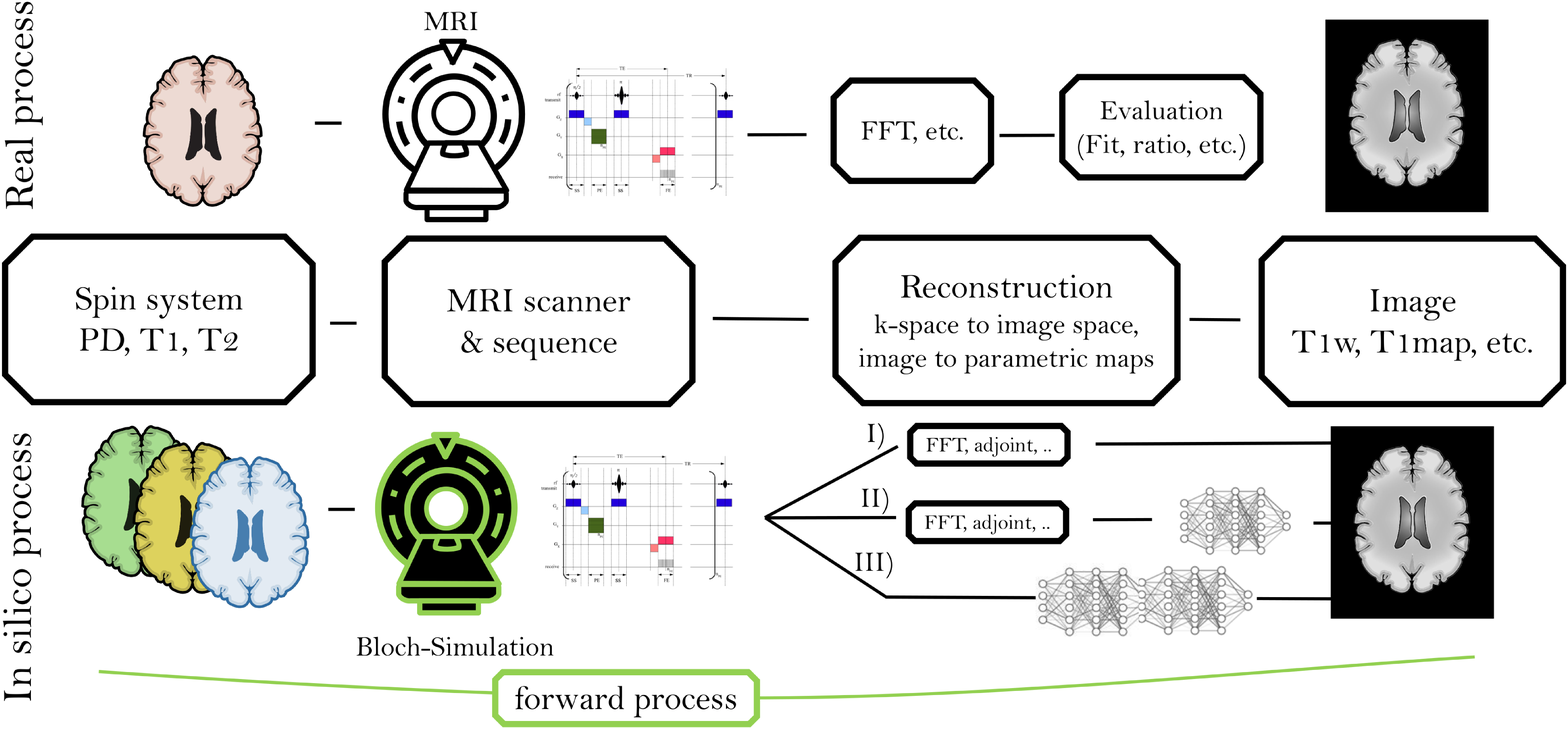}
\caption{\textit{MRzero} differentiable forward process. The general MRI pipeline from spin system to reconstructed image. For this pipeline a differentiable MR scanner simulation implements Bloch equations for signal generation, and for reconstruction either I) conventional reconstruction, II) conventional reconstruction with a neural network appended, or III) a CNN/NN reconstruction is used. Although path III is envisioned, it is not implemented in current work. This differentiable forward process is a fundamental building block of the MR zero flow scheme shown in Figure 2.}
\end{figure}

\begin{figure}[!ht]
\centering
\includegraphics[width=14cm]{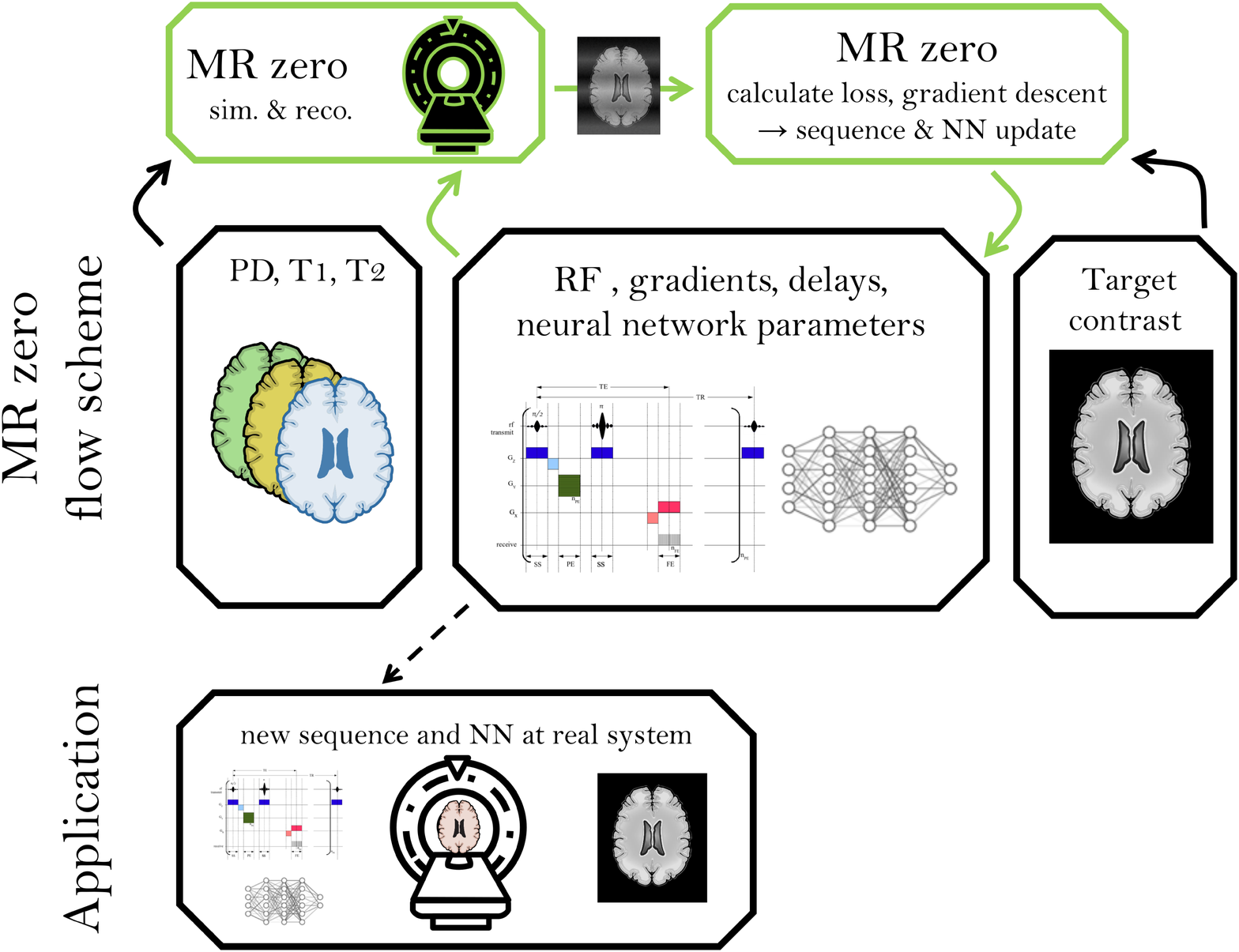}
\caption{\textit{MRzero} flow schematic. The output of \textit{MRzero} forward process calculation is compared to the target image, analytical derivatives w.r.t. sequence parameters are computed using auto-differentiation, and gradient descent is performed in parameter space to update both sequence and NN parameters. This whole process is repeated until a convergence criterion is satisfied subject to the loss function imposed on parameters/outputs. Application: Final or intermediate sequences can then be applied at the real scanner using the pulseq framework. \cite{layton_pulseq_2017}}
\end{figure}

\begin{figure}[!ht]
\centering
\includegraphics[width=14cm]{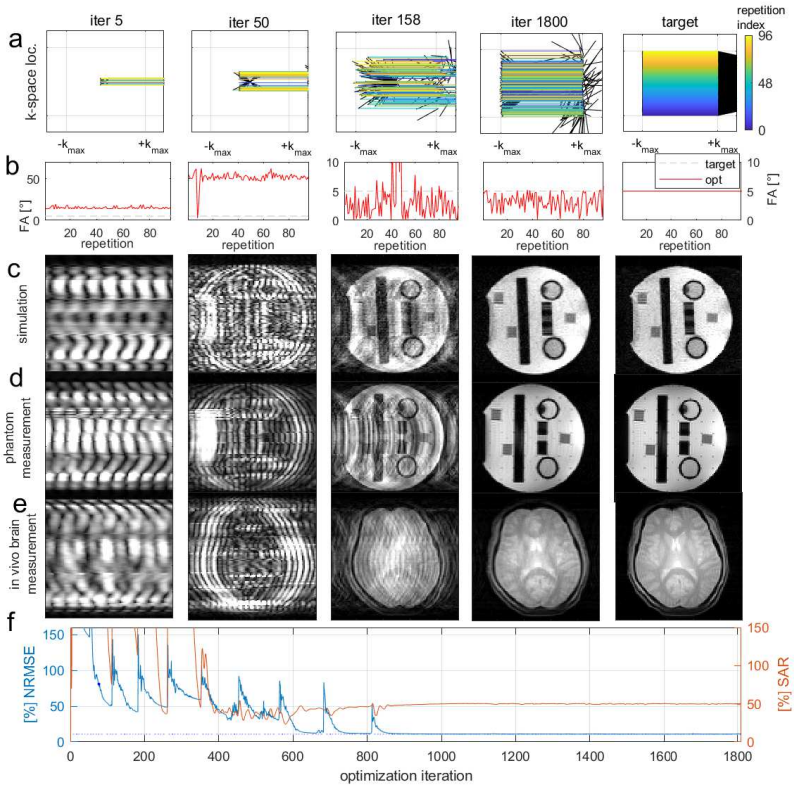}
\caption{RF and encoding task. (a) k-space sampling at different iterations. (b) Flip angles over measurement repetitions. (c) Simulation-based reconstruction at different iterations 9, 99, 255, 355, and 1000. (d) Phantom measurement. (e) In vivo brain scan. (f) Training error curve. An animated version can be found at: www.tinyurl.com/y4blmpe7. Target sequence: 2D transient gradient- and RF-spoiled GRE, matrix size 96, TR = 25 ms, TE = 3.2 ms, FA=5$^{\circ}$.}
\end{figure}

\begin{figure}[!ht]
\centering
\includegraphics[width=13cm]{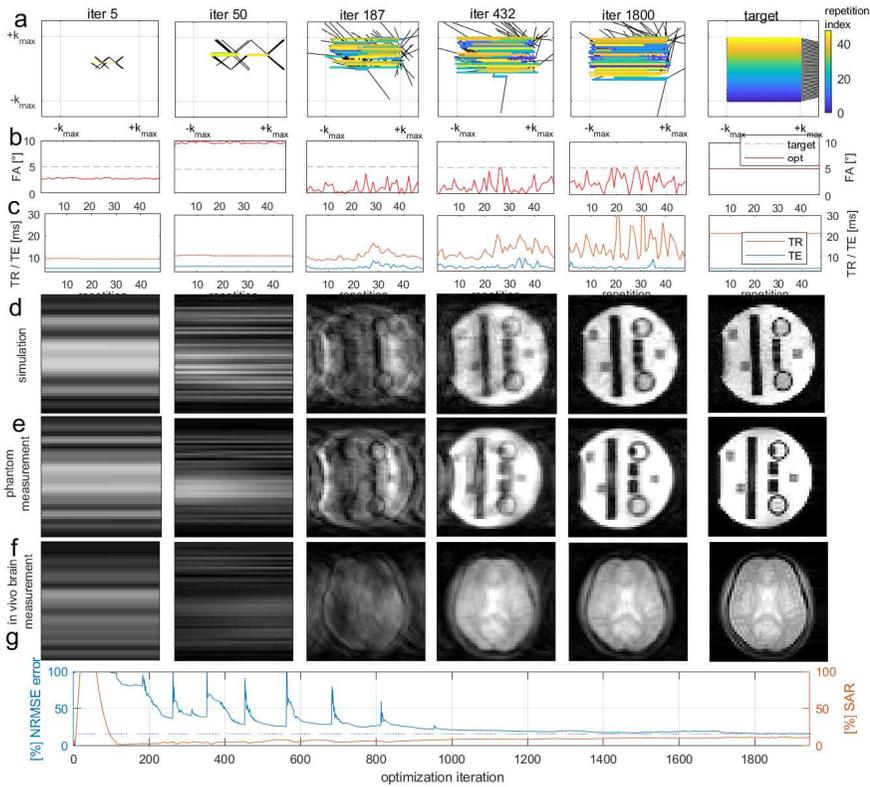}
\caption{RF, Encoding and timing task. (a) k-space sampling at different iterations. (b) Flip angles over measurement repetitions. (c) TR and TE over measurement repetitions. (d) Simulation-based reconstruction at different iterations. (e) Phantom measurement, (f) In vivo brain scan. (g) Training error curve. An animated version can be found at: www.tinyurl.com/y4l4mrrv. Target sequence: 2D transient gradient- and RF-spoiled GRE, matrix size 48, TR = 20 ms, TE = 3 ms, FA=5$^{\circ}$.}
\end{figure}

\begin{figure}[!ht]
\centering
\includegraphics[width=13cm]{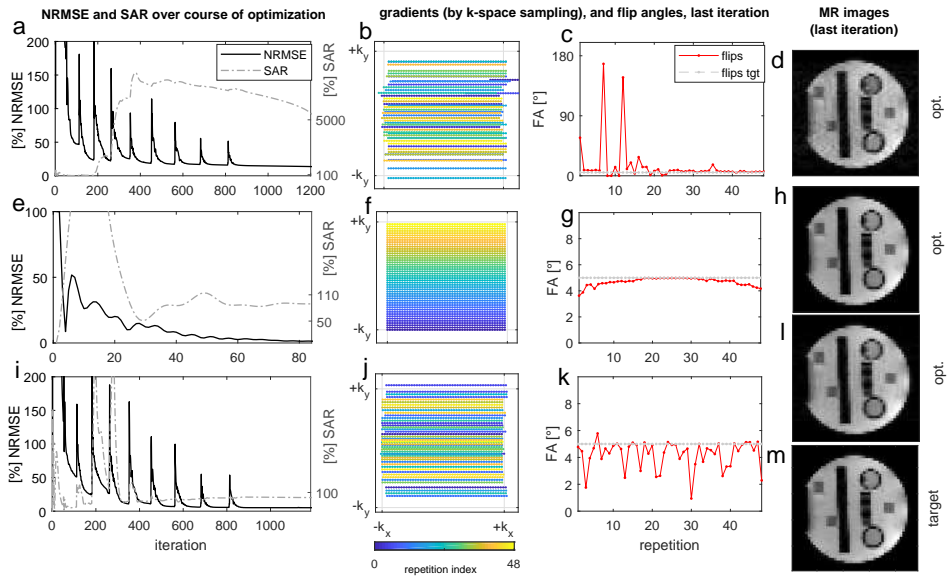}
\caption{SAR regularization. First row: \textit{Encoding and RF learning task} with no SAR penalty, which leads to a solution with high flip angles. Second row: \textit{RF optimization} for linear reordered gradient encoding using a SAR penalty that regularizes only the flip angle amplitudes. Third row: \textit{Encoding and RF learning task} as done in Figure 2 with gradient and RF optimization upon SAR penalty. (a,e,i) Training error curves and relative SAR over optimization steps. (b,f,j) k-space sampling at the last iteration. (c,g,k) Flip angles at the last iteration (red) compared to target flip angles (gray). (d,h,l) Optimized sequence image and target image.  Target sequence: 2D transient gradient- and RF-spoiled GRE, matrix size 48, TR = 20 ms, TE = 1.5 ms, FA=5$^{\circ}$.}
\end{figure}

\begin{figure}[!ht]
\centering
\includegraphics[width=14cm]{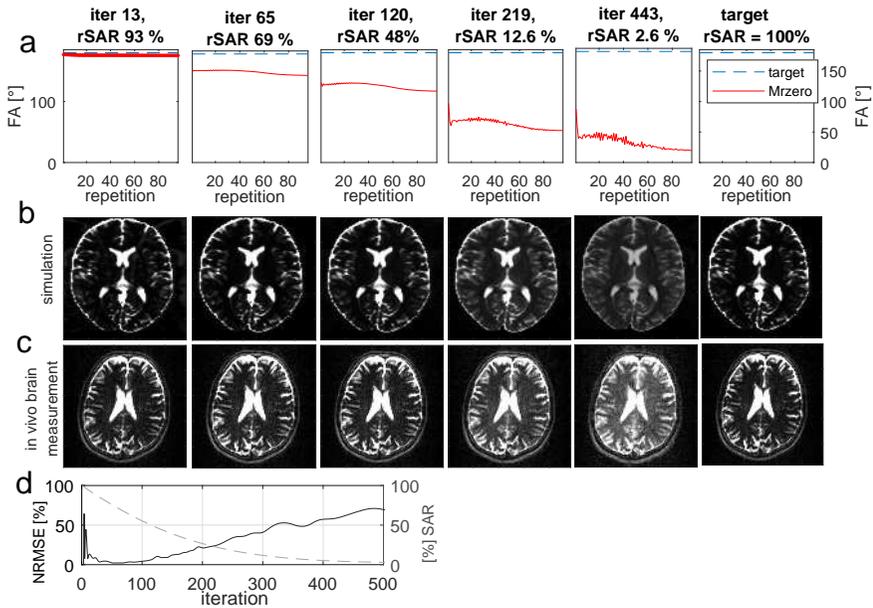}
\caption{RARE sequence trained on numerical brain phantom. (a) Flip angles at at different iterations. (b) Simulation-based reconstruction at iterations 13, 65, 120, 291 and 443. (c) In vivo brain scan. (d) Training error curve and SAR relative to the target.
Target sequence: single-shot RARE, matrix size 96, linear reordering with k-space center at rep. 48, TA = 0.59 s, echo spacing = 6 ms, excitation FA=90$^{\circ}$, refocusing FA=180$^{\circ}$.}
\end{figure}

\begin{figure}[!ht]
\centering
\includegraphics[width=13cm]{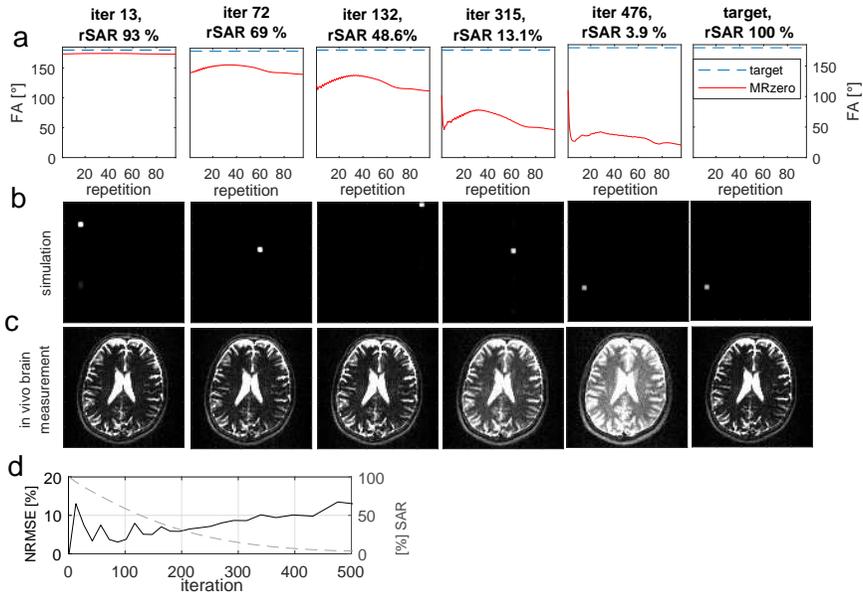}
\caption{ 
RARE sequence trained on a sparse synthetic database. (a) flip angles at at different iterations. (b) Simulation-based reconstruction at iterations 13, 72, 132, 318 and 476. The apparent change of position of the voxel is caused by the fact that at each iteration a random sample from the dataset is used, and we use just a single sample to show how the target looks. (c) In vivo brain scan. 
(d) Training error curve and SAR relative to the target for the sparse synthetic dataset approach. Target sequence: single-shot RARE, matrix size 96, linear reordering with k-space center at rep. 48, TA = 0.59 s, echo spacing = 6 ms, excitation FA=90$^{\circ}$, refocusing FA=180$^{\circ}$. An animated version can be found at: https://tinyurl.com/y9w899uv}
\end{figure}

\begin{figure}[!ht]
\centering
\includegraphics[width=14cm]{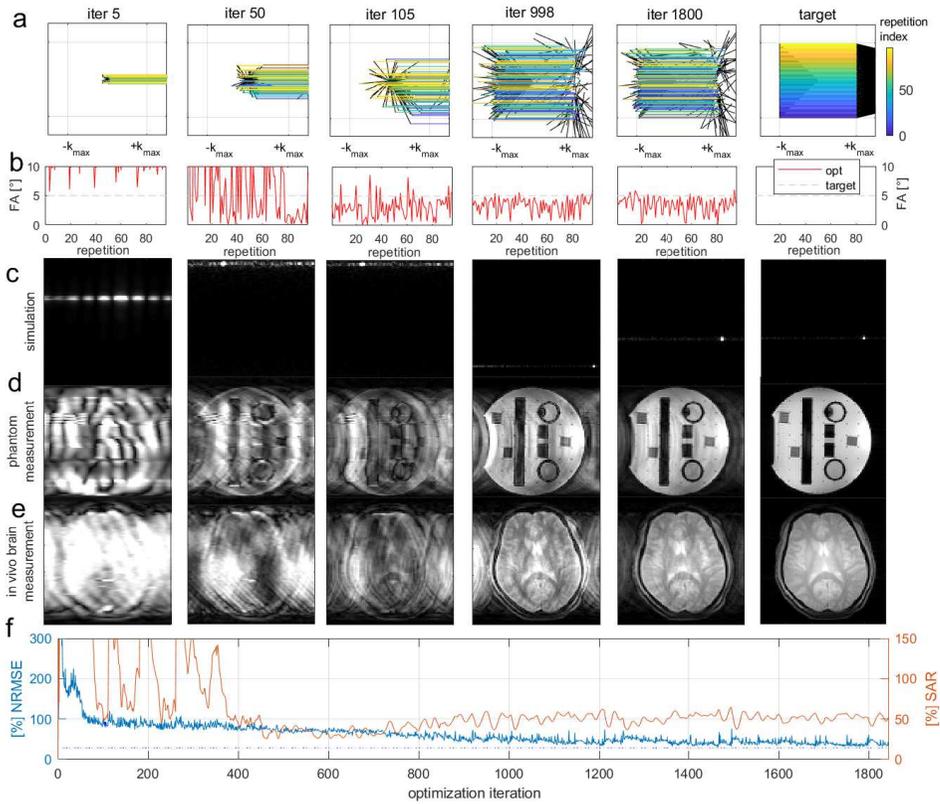}
\caption{GRE sequence trained on a sparse synthetic database (a) k-space sampling at different iterations. (b) flip angles at different iterations. (c) Simulation-based reconstruction (trained on sparse synthetic isochromats) at iterations 111, 551, 991, 2091, and 3191. The apparent change of position of the voxel is caused by the fact that at each iteration a random sample from the dataset is used, and we use just a single sample to show how the target looks. (d) Phantom measurement. (e) In-vivo measurement. (f) Training error curve. Target sequence: 2D transient gradient- and RF-spoiled GRE, matrix size 96, TR = 25 ms, TE = 3.2 ms, FA=5$^{\circ}$. Animated version can be found at: https://tinyurl.com/y6pzmxq7}
\end{figure}

\begin{figure}[!ht]
\centering
\includegraphics[width=14cm]{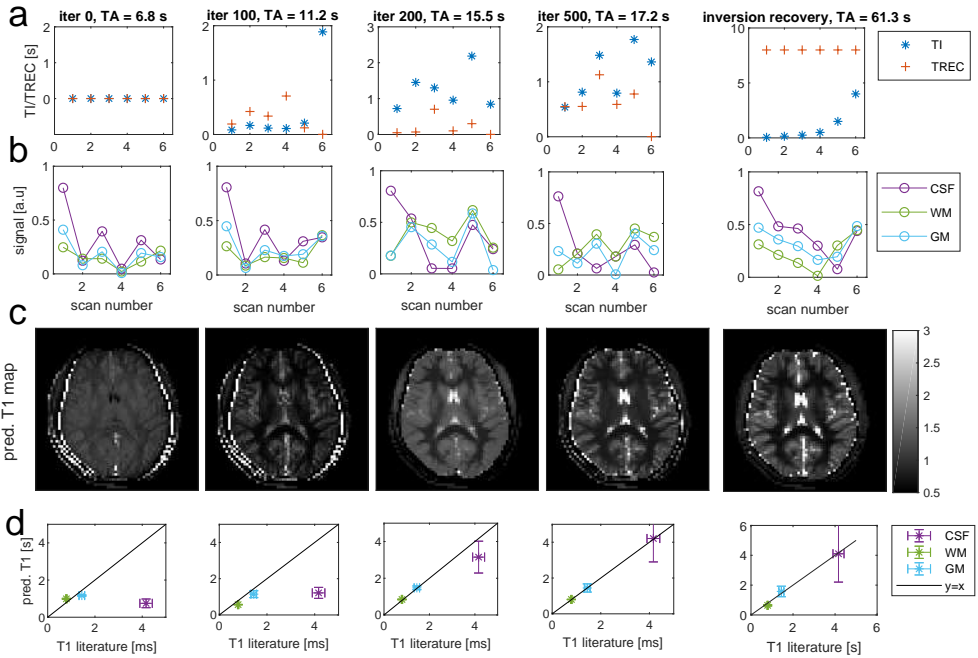}
\caption{Auto-encoding of T1. Strategies here were found using the sparse synthetic database approach, but all images are real in vivo scans. (a) Different stages of the sequence while optimizing TI and Trec times, which were initialized to 0. (b) Corresponding signal evolution for different ROIs. (c) The NN reconstruction predicts the T1 in each pixel. The initial step is strongly biased, but after several iterations \textit{MRzero} adjusts both relaxation and inversion times until it's able to generate a T1map that matches well to literature values at 3 T (d). The rightmost column shows a standard inversion recovery sequence with typical inversion times for reference. Signals of this sequence were also evaluated using an adapted neural network. Sequence details: 180 degree inversion prepared 2D transient gradient- and RF-spoiled centric reordered GRE, matrix size 64, TR = 15 ms, TE = 8 ms, FA=5$^{\circ}$ repeated for inversion 6 times. Animated version can be found at: www.tinyurl.com/y9qe7b7m . A higher resolution version can be found in Supporting Information Figure S7}
\end{figure}

\end{document}